\documentclass[aps,pra,twocolumn,groupedaddress]{revtex4-1}
\usepackage{graphics,graphicx,float,amsmath, xcolor} 
\usepackage{upgreek}
\usepackage{hyperref}
\usepackage{hhline}

\begin{document}

\title{Charge State Dynamics During Excitation and Depletion of the Nitrogen Vacancy Center in Diamond}
\author{Luke Hacquebard}
\affiliation{McGill University}
\author{Lilian Childress}
\affiliation{McGill University}
\date{\today}
\begin{abstract}
	The charge state dynamics of the nitrogen-vacancy (NV) center in diamond play a key role in a wide range of applications, yet remain imperfectly understood. Using single ps-pulses and pulse pairs, we quantitatively investigate the charge dynamics associated with excitation and fluorescence depletion of a single NV center. Our pulsed excitation approach permits significant modeling simplifications, and allows us to extract relative rates of excitation, stimulated emission, ionization, and recombination under 531 nm and 766 nm illumination. By varying the duration between paired pulses, we can also investigate ionization and recombination out of metastable states. Our results are directly applicable to experiments employing stimulated emission-depletion imaging, and can be used to predict optimal operating regimes where excitation and stimulated emission are maximized relative to charge-state-switching processes.
\end{abstract}
\maketitle

\section{Introduction}

The nitrogen-vacancy (NV) center in diamond has become a workhorse for quantum information and metrology~\cite{doherty_nitrogen-vacancy_2013, schirhagl_nitrogen-vacancy_2014}, with applications ranging from quantum networks~\cite{nemoto_photonic_2014} to geology~\cite{glenn_micrometerscale_2017} to fundamental spin-bath physics~\cite{hanson_coherent_2008}. Such applications rely on the favorable spin properties of the negatively charged state (NV$^-$), which exhibits long coherence times and permits optical preparation and detection~\cite{doherty_nitrogen-vacancy_2013}. At the same time, controlled conversion to the neutral charge state (NV$^0$) offers opportunities for optical nanoscopy~\cite{han_metastable_2010, chen_near-infrared-enhanced_2017}, charge-based memories~\cite{dhomkar_long-term_2016, jayakumar_optical_2016}, electrical spin detection~\cite{bourgeois_photoelectric_2015, hrubesch_efficient_2017}, and improved spin detection via spin-to-charge conversion~\cite{shields_efficient_2015, hopper_near-infrared-assisted_2016, jaskula_improved_2017, hopper_amplified_2017}. Understanding and controlling the charge state of the NV thus underlies a broad range of potential technologies, yet the dynamics of optically-induced charge state switching remains an area of active research~\cite{beha_optimum_2012, aslam_photo-induced_2013, siyushev_optically_2013, chen_spin_2015, berthel_photophysics_2015, ji_charge_2016}.

%At the same time, intense optical pulses can be used to control the internal states of the NV$^-$ defect, both coherently~\cite{bassett_ultrafast_2014} and incoherently, where pulses can induce both optical excitation and depletion of fluorescence.
In this paper, we focus on quantitative measurements of charge state switching associated with pulsed optical excitation and depletion processes, where a pulse of green light excites the NV and a pulse of red or near-IR light causes stimulated emission. Such two-color pulsed illumination is employed in stimulated emission-depletion (STED) imaging, permitting imaging of NV centers with nanoscale resolution~\cite{rittweger_sted_2009, wildanger_diffraction_2011, wildanger_solid_2012}. Understanding the associated charge state dynamics provides further insight into the fluorescence depletion mechanism -- stimulated emission or ionization -- which also has implications for proposed laser threshold magnetometry~\cite{jeske_laser_2016}. % -- which has recently been studied qualitatively in defect ensembles~\cite{jeske_stimulated_2017}. 
Moreover, if it is possible to induce excitation and stimulated emission of NV$^-$ without ionization, then several  opportunities open up. Controlled transfer into and out of the NV$^-$ excited state could enable fast electron-nuclear spin gates~\cite{fuchs_excited-state_2010}.  Alternately, if STED does not ionize and preserves the polarization~\cite{wildanger_diffraction_2011} or coherence of surrounding spins, it may become possible to detect the joint state of several closely-spaced, interacting spins~\cite{dolde_high-fidelity_2014}. 

Our experiments examine ionization (NV$^- \rightarrow$ NV$^0$) and recombination (NV$^0 \rightarrow$ NV$^-$) induced by sequential pulses of green (531 nm) and red (766 nm) light, at wavelengths commonly used for STED experiments. Essentially, we ask the questions: (1) can we directly compare the excitation rate of green light to rates of ionization/recombination out of the excited states? (2) how much does red light cause stimulated emission vs ionization/recombination? (3) how does red-induced ionization out of the NV$^-$ metastable singlet state compare to ionization out of the excited state? 

We study single NV centers, where we can prepare and detect the charge state with high fidelity~\cite{aslam_photo-induced_2013, shields_efficient_2015} and thereby determine ionization and recombination probabilities quantitatively. In contrast to previous work that examined longer-timescale dynamics of green and near-IR illumination~\cite{hopper_near-infrared-assisted_2016, ji_charge_2016}, we examine single $\sim$100 ps-pulses or pulse pairs. %Since we control the initial charge and spin configuration and apply only single pulses or pulse-pairs, we can use simplified level structures to model the system with linear rate equations.  
Working in this single-pulse regime eliminates long-time-scale dynamics associated with spin and metastable states, and reveals a simple picture of optically-induced processes. We model the system with linear rate equations that we quantitatively fit for excitation, stimulated emission, ionization, and recombination rates. The model can then be used to predict optimal operating regimes and behaviors.  We note that our approach of linear rate equations does not contradict the quadratic dependence of ionization on power observed at low intensities in e.g.~\cite{aslam_photo-induced_2013}. Because we explicitly include the excited states in our model (the $^3$E excited state of NV$^-$ and the $^2$A$_1$ state of NV$^0$~\cite{doherty_nitrogen-vacancy_2013, gali_theory_2009}), each process (excitation followed by ionization) can be linear while the overall ionization process still requires two photons. %At first glance, some of our experimental measurements appear to contradict earlier results on simultaneous green and near-IR illumination~\cite{hopper_near-infrared-assisted_2016, ji_charge_2016}, %These results complement existing models for near-IR behavior (at 900-1000 nm ~\cite{hopper_near-infrared-assisted_2016} and 1064 nm~\cite{ji_charge_2016}), as well as 

For example, our results corroborate ensemble measurements indicating that stimulated emission dominates over ionization~\cite{jeske_stimulated_2017}, and additionally provide quantitative values that will make it %: 766 nm-induced fluorescence depletion from the NV$^-$ excited state is associated with simulated emission 93.3$\pm$0.2\% and ionization 6.6$\pm 0.2$\% of the time, making 
it possible to more accurately model potential NV lasing behaviour. 
We also measure the rate of ionization by 766 nm light out of the NV$^-$ excited and singlet states, which may inform spin-to-charge conversion approaches that utilize near-IR singlet ionization~\cite{hopper_near-infrared-assisted_2016}. Perhaps disappointingly, our results indicate that excitation and depletion are inevitably accompanied by ionization, and at best an ``ideal" excitation-stimulated emission cycle occurs with 82.5$\pm$0.3\% probability. Our model and extracted rates thus provide a framework for understanding and designing a wide range of experiments. %model can be used to predict optimal operating regimes.   %At the same time, optical control over the internal states of the NV$^-$ charge state 

%\begin{figure}
%\includegraphics{SGP_Ionization.pdf}%
%\caption{\label{Two}}
%\end{figure}

\section{Experimental methods}

Our experiments employ four optical excitation pathways, allowing charge state preparation, manipulation, and detection for a single NV in bulk diamond.  Briefly, continuous-wave (CW) green (532 nm) excitation is used to initialize the distribution of NV charge and spin populations at the beginning of the experiment, while single-shot charge state preparation and measurement is performed using yellow light (594 nm) \cite{aslam_photo-induced_2013}. We thereby extract the effects of picosecond pulses of green (531 nm) and red (766 nm) excitation on the NV charge state. Additional microwave control over the NV spin permits us to examine spin-dependence of the charge dynamics.

More specifically, our laser sources comprise CW 532 nm (LaserQuantum Ventus), CW 594 nm (Newport R-39582 HeNe), pulsed 531 nm and pulsed 766 nm lasers (both PicoQuant LDH-FA, with pulse widths $<$ 100 ps); each source is gated by an acousto-optic modulator (AOM). For simplicity, we will refer to the 532/531 nm lasers as ``green", the 594 nm laser as ``yellow"€™ and the 766 nm laser as ``red™." The two pulsed lasers are also separately controlled by a SEPIA II laser driver (PicoQuant PDL 828), allowing for tuning of %the repetition rate and 
pulse separation timing. The four optical excitation paths are combined in a homebuilt confocal microscope, %(using either dichroic mirrors or polarizing beamsplitters) 
and focused through a 1.35 NA oil objective (Olympus UPlanSApo 60x) onto a single NV in a type IIa $\langle 111 \rangle$-cut, chemical-vapor-deposition-grown diamond sample (Sumitomo). NV center fluorescence in the wavelength range 635 - 750 nm is detected with %collected into a single mode optical fiber, and directed to 
a single photon counting module (Picoquant $\tau$-SPAD). The sample is mounted on an XYZ scanning piezo stage (nPoint NPXY100Z25-219), and a 20 micron copper wire soldered across the diamond allows application of microwave signals for driving spin transitions. Except for the pulsed laser driver, the timing of the experiments is determined by a field programmable gate array (FPGA) card (National Instruments 7841-R FPGA) that controls the scanning microscope and records photon counts; the FPGA card also outputs digital pulse patterns to rapidly turn on and off the laser AOMs %the acousto-optic modulators 
and microwaves with 8.3 ns resolution \cite{ziegler_notitle_2012, gupta_efficient_2016}. 

Most of our experiments use a pulse sequence (see e.g. Fig.~\ref{Fig1}a) that incorporates a spin and charge preparation step before pulsed illumination, and a charge detection step afterwards. %The generic pulse sequence used in our experiments is shown in Fig.~\ref{Fig1}a. 
CW green illumination (2 $\upmu$s, $\sim$250 $\upmu$W) is used to fix the initial charge state distribution to approximately 70$\%$ NV$^-$, and also initializes the NV$^-$ spin primarily into $m_s = 0$. Long, low power CW yellow pulses (4 ms, 2 $\upmu$W) are used to initialize and read out the NV center's charge state before and after pulsed illumination. Yellow illumination efficiently excites NV$^-$, while only weakly exciting NV$^0$, leading to a high fluorescence contrast between the two charge states \cite{aslam_photo-induced_2013}. The number of photon counts collected during a yellow pulse can be used to determine the initial and final charge states of that pulse~\cite{shields_efficient_2015}, with fidelities in the range 91-97$\%$ in our experiments. These measurements allow us to extract the probability that the applied pulsed illumination caused ionization (NV$^- \rightarrow $NV$^0$) or recombination (NV$^0 \rightarrow$ NV$^-$); see appendix~\ref{AppendixA} for more information. In order to investigate single pulses we fix the pulsed lasers' repetition rate to 1 MHz and turn on their acousto-optical modulators for a 1 $\upmu$s illumination window so that on average only single pulses will reach the sample. Fluorescence counts are also collected during this 1 $\upmu$s window, revealing charge-dependent fluorescence saturation behaviour. %Illumination with an exact number of pulses has not yet been realized due to the complexities involved with synchronizing the pulsed laser output with the FPGA system. 

We acquire data with three types of measurements that give us information about different aspects of the system dynamics. (1) Charge state switching data (obtained using photon counts during yellow illumination) reveal ionization and recombination processes; (2) Fluorescence counts (obtained during pulsed green illumination) provide information on the efficiency with which internal states are excited; (3) Time-resolved fluorescence detection allows us to examine fluorescence depletion processes induced by red illumination. Due to slow data acquisition rates, this third measurement is performed with only green and red pulsed excitation (without CW green or yellow illumination, see Fig.~\ref{Fig2}a). By simultaneously fitting all three types of measurement to a rate equation model, we can well constrain the rates of excitation, stimulated emission, ionization, and recombination under both red and green illumination. 

\section{Pulsed green excitation and charge-state switching}

\begin{figure}
	\includegraphics[scale=0.23]{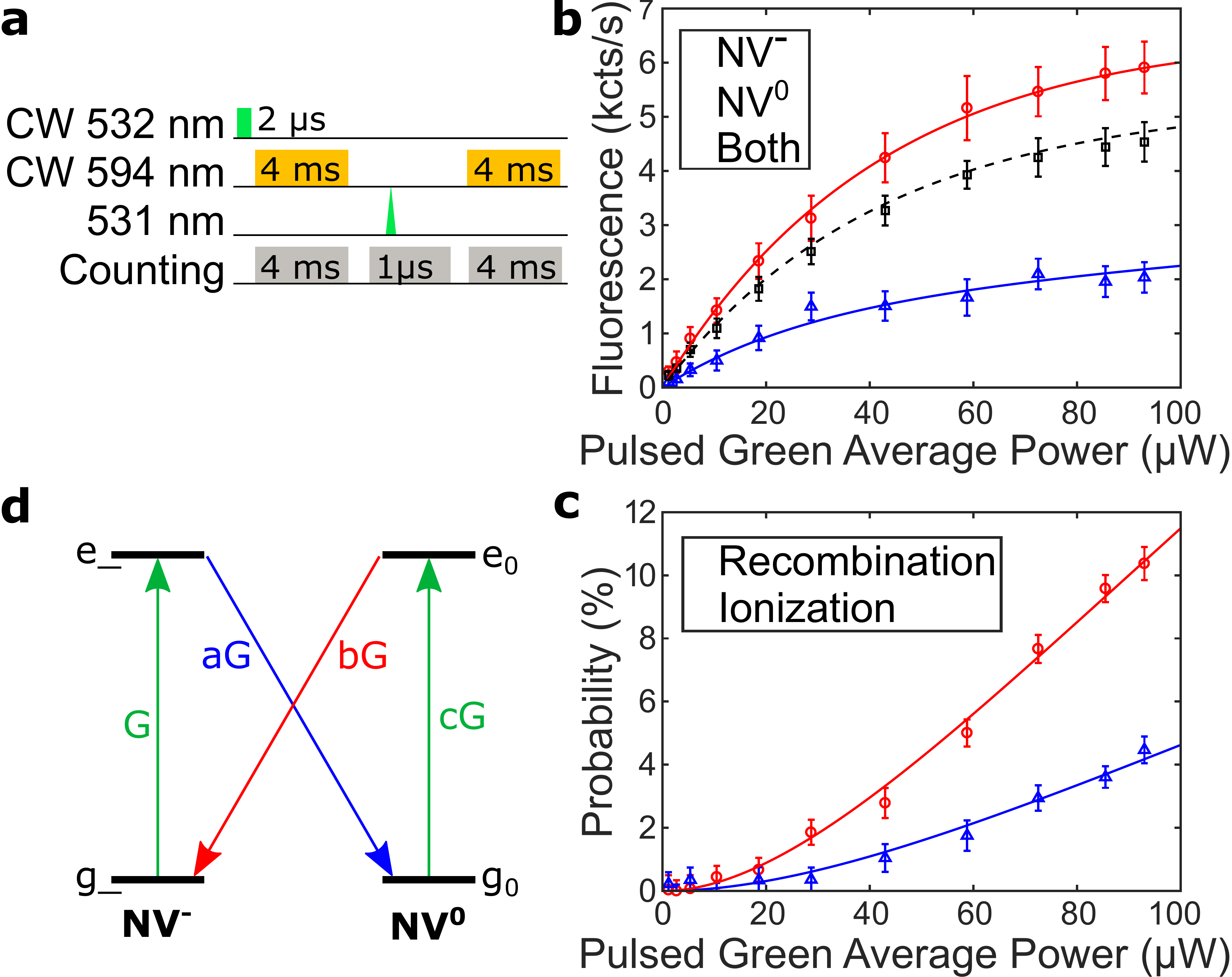}%
	\caption{\label{Fig1}Excitation and charge state switching under green illumination. \textbf{(a)} Pulse sequence used for data shown in (b) and (c).  CW green (532 nm)  is used to initialize the charge state distribution; low-power CW yellow (594 nm) is used to measure the NV charge state before and after pulsed illumination. Between the yellow pulses, a green pulse is applied. \textbf{(b)} Charge-dependent fluorescence during pulsed green excitation as a function of pulsed green average power.  Fluorescence for NV$^-$ initialization (red circles) is larger than for NV$^0$ initialization (blue triangles) primarily due to spectral filtering. Black squares show total fluorescence without charge selection. \textbf{(c)} Single green pulse probability of ionization (blue triangles) and recombination (red circles). \textbf{(d)} Four-level rate equation model including transitions between the ground $g_- (g_0)$ and excited states $e_- (e_0)$ of NV$^-$ and NV$^0$ respectively. Here we show the transition rates under green illumination. Solid lines in \textbf{b} and \textbf{c} are found by using this four-level model to simultaneously fit data shown in Figs.~\ref{Fig1} and \ref{Fig2}.}
\end{figure}

Pulsed green illumination can promote both the negative and neutral charge states of the NV center to an excited state, and can cause subsequent ionization or recombination to occur~\cite{aslam_photo-induced_2013, siyushev_optically_2013}. Efficient excitation with minimal ionization is often desired when studying NV$^-$, as any switching to NV$^0$ will result in reduced signal-to-noise and loss of spin polarization \cite{chen_spin_2015}. We investigate the probabilities of excitation, ionization and recombination of single green pulses using the pulse sequence shown in Fig.~\ref{Fig1}a, while recording both fluorescence and charge state data. 

Using the first yellow pulse to select the initial charge state, we measure the charge-dependent fluorescence as a function of pulsed green average power (with the AOM enabled), shown in Fig.~\ref{Fig1}b. Data sets with high photon counts ($>$ threshold = 1)
during the first yellow pulse are assigned to NV$^-$, while those with low counts are assigned to NV$^0$~\cite{aslam_photo-induced_2013, shields_efficient_2015}. Because our spectral filters are optimized for NV$^-$ emission, we obtain a higher fluorescence rate for NV$^-$ initialization (red circles) than NV$^0$ initialization (blue triangles). We also show the non-conditional total measured fluorescence of both charge states (black squares) resulting from the overall initial $63.8\pm0.2\%$ NV$^-$ population. In the simultaneous fits described below, these data sets most tightly constrain the excitation probability of the green pulse within the NV$^-$ and NV$^0$ manifolds. 

The probability of ionization and recombination as a function of green power (see Fig.~\ref{Fig1}c) is found by comparing the charge state measured before and after the green pulse, and accounting for the charge state readout fidelities (see appendix~\ref{AppendixA}). Recombination (red circles) dominates over ionization (blue triangles), which is expected because green illumination preferentially populates NV$^-$. %While we observe a quadratic dependence at low powers, where two photons are required to first excite and then ionize/recombine~\cite{aslam_photo-induced_2013}, the approximately linear increase in probability at higher powers is likely due to saturation of the excited states, leading to a one-photon process for ionization and recombination out of the already-occupied excited state. 
We observe a quadratic dependence at low powers, where the instantaneous intensity is below the saturation intensity of the optical transitions, such that two photons are required to first excite and then ionize/recombine~\cite{aslam_photo-induced_2013}. The approximately linear increase in probability at higher powers is likely due to saturation of the excited states, leading to a one-photon process for ionization and recombination from the excited states.
Note that our measurement scheme only considers the initial and final charge state; thus at even higher powers, these probabilities should saturate at values $<100\%$ due to the possibility of multiple switching events occurring within a single pulse.

%One thing to note here is that our measurement is not a measure of a single ionization event but instead a measure of ending up in the opposite charge state, leading to the possibility of two charge state switching events occuring within the pulse. This means our probabilities will never reach 100 percent...

%(By investigating the single green pulse probabilities of excitation, ionization, and recombination we can develop a model of the NV center system in order to predict the optimum power for maximizing excitation while minimizing ionization. )

The solid lines shown in Fig.~\ref{Fig1}b/c are results of a four-level rate equation model %(see Figs.~\ref{Fig1}d and ~\ref{Fig2}d)
 simultaneously fit to data in Fig.~\ref{Fig1} and \ref{Fig2}. %; the curves do not change perceptibly if we fit only to data in Fig.~\ref{Fig1}. 
 Since the data in Fig.~\ref{Fig1} involves only green pulses, it can be modeled using the transition rates shown in Fig.~\ref{Fig1}d. The model comprises the ground and excited states of NV$^-$ and NV$^0$, with the relevant rates shown for green-induced excitation ($G$ for NV$^-$ and $c G$ for NV$^0$), ionization ($a G$) and recombination ($b G$). Note that all rates are linear in the green power $\propto G$; because we explicitly include excited states in our model, it does not contradict nonlinear models that do not include excited states~\cite{hopper_near-infrared-assisted_2016,aslam_photo-induced_2013}. This model corresponds to proposed physical mechanisms for ionization (via excitation to the conduction band and Auger recombination) and recombination (via excitation from the valence band)~\cite{siyushev_optically_2013}\footnote{There is not a clear consensus on where the system ends up after ionization out of NV$^-$, with some papers proposing that the NV$^0$ excited state can be populated~\cite{ji_multiple-photon_2017, aslam_photo-induced_2013}. We also considered this possibility, and fit our data to an alternate model where ionization processes (both red and green) leave the system in the NV$^0$ excited state rather than ground state. We found that the resulting fit parameters associated with green excitation, ionization and recombination all lie within error of previously quoted results. The parameters associated with red-induced stimulated emission, ionization and recombination were found to shift slightly to $d=9.7\pm0.4\%$, $e=25\pm2\%$, and $f=84\pm7\%$. The resulting weighted sum of squared errors when ionizing to the NV$^0$ ground state versus the NV$^0$ excited state were found to be 18.5 and 24.1 respectively, indicating that for our data a model consisting of ionization to the NV$^0$ ground state is more likely to be correct.}.  
 Note that the NV$^-$ singlet state is disregarded because it is populated on timescales ($\sim 10-100$ ns~\cite{gupta_efficient_2016}) much longer than the pulse widths ($\sim 100$ ps) and pulse separations ($\sim 0.6$ ns) relevant to the measurements in Figs.~\ref{Fig1} and \ref{Fig2}. %Since the data in Fig.~\ref{Fig1}b/c involves only green excitation, its behaviour
%We simultaneously model and fit the data shown in Figs.~\ref{Fig1} and \ref{Fig2}, as discussed below; the curves shown in Fig.~\ref{Fig1} are the results of this simultaneous fit.
%This four-level model will also be used for fitting the red-dependent data (discussed below) and the fits to the green- (Fig.~\ref{Fig1}) and red- (Fig.~\ref{Fig2}) dependent data sets are performed simultaneously to allow for proper error analysis. In both cases the NV$^-$ singlet state can be ignored, as the timescale for it to be populated is much longer than the pulse widths ($\sim 100$ ps) and pulse separation time ($\sim 0.6$ ns) in these experiments. 
Similarly, spontaneous emission can be neglected during the short optical pulses. We can also ignore dependence on the NV$^-$ spin states in our model because the spin is initialized each time by the same CW green pulse and yellow pulse, spin-dependent singlet transitions are too slow to affect the dynamics, and ionization rates out of the excited state are measured to be spin-independent (data not shown). Similarly, we expect any slow processes within the NV$^0$ state to be negligible with the consistent initialization procedure and on the sub-nanosecond timescales we study. It is worth emphasizing that this considerable simplification in modeling is obtained by working with single pulses, and ensuring that the charge and spin are identically initialized before each pulse.

We model the single-green-pulse, charge-dependent fluorescence data (Fig.~\ref{Fig1}b) by calculating the NV$^-$ and NV$^0$ excited state populations after the green pulse for each initialization condition. In particular, we start the model in a combination of NV$^-$ and NV$^0$ ground states with a charge state distribution set by the initialization fidelities of the first yellow pulse (specifically, to model ``NV$^-$" fluorescence, the initial NV$^-$ population is $91.25\pm0.2\%$ and the NV$^0$ population is $8.75\pm0.2\%$, whereas for modeling ``NV$^0$" fluorescence, the NV$^-$ population is $5.0\pm0.2\%$ and the NV$^0$ population is $95.0\pm0.2\%$). We then allow the model (with rates shown in Fig.~\ref{Fig1}d) to evolve for 100 ps to simulate the green pulse. The total fluorescence from the pulse is then given by $\alpha_- e_-(t) + \alpha_0 e_0(t)$, where $e_{-}(t)$ ($e_{0}(t)$) is the excited state NV$^-$ (NV$^0$) population after the pulse and $\alpha_-$ ($\alpha_0$) is a scaling factor comprising the collection efficiency and quantum efficiency for NV$^-$ (NV$^0$) emission. 
%The final excited state populations of each charge state are each multiplied by their own fit parameter ($\alpha_-$ or $\alpha_0$ for NV$^-$ or NV$^0$ respectively, corresponding to a fluorescence scaling factor comprising collection efficiency and quantum efficiency) and added together to model the measured fluorescence. 
The dashed black line in Fig.~\ref{Fig1}b corresponds to a model prediction of fluorescence given an initial 63.8$\%$ NV$^-$ population. We find the ionization and recombination probabilities (Fig.~\ref{Fig1}c, solid lines) by initializing the model entirely into either ground state (since we already correct the probabilities in Fig.~\ref{Fig1}c for imperfect charge-state initialization) and calculating the population of the opposite charge state after 100 ps of green illumination.

%The four-level model fits for the single-green-pulse, charge-dependent fluorescence data (Fig.~\ref{Fig1}b, solid lines) are performed by first initializing into the ground states of NV$^-$ and NV$^0$ with a charge state distribution set by the initialization fidelities of the first yellow pulse. The measured initialization fidelities results in an initial NV$^-$ population of $91.25\pm0.2\%$ for modeling NV$^-$ fluorescence and $5.0\pm0.2\%$ for modeling NV$^0$ fluorescence. After ground state initialization, we apply the green-induced rates shown in Fig.~\ref{Fig1}d for 100 ps. The final excited state populations of each charge state are then multiplied by their own fit parameter, corresponding to a fluorescence scaling factor, and added together to give the final measured fluorescence. The dashed black line in Fig.~\ref{Fig1}b corresponds to a model prediction of fluorescence given an initial 63.8$\%$ NV$^-$ population. The fits for ionization and recombination (Fig.~\ref{Fig1}c, solid lines) are made by initializing entirely into either ground state and looking at the probability to end up in the opposite charge state after a 100 ps pulse of green illumination. 

The fits shown in Fig.~\ref{Fig1} are made simultaneously with other data sets (see discussion of Fig.~\ref{Fig2}). In this simultaneous fit, the data shown in Fig.~\ref{Fig1} provide the tightest constraints on the rates for green-induced excitation, ionization and recombination. Since the exact pulse width is unknown, we quote our results in terms of a percentage of the NV$^-$ excitation rate ($G$), as shown in Fig.~\ref{Fig1}d. We obtain relative rates of green-induced ionization $a=3.7\pm0.6\%$, recombination $b=8\pm1\%$, and NV$^0$ excitation $c=130\pm20\%$. Using the two fluorescence scaling factors extracted from the fit, along with an NV$^-$ excited state lifetime of $12.2\pm0.1$ ns (when optically pumped into $m_s = 0$, measured in Fig.~\ref{Fig2}a), and an NV$^0$ excited state lifetime of $18\pm3$ ns (taken to overlap literature values \cite{beha_optimum_2012, liaugaudas_luminescence_2012, berthel_photophysics_2015}), we calculate that $15\pm2\%$ of our collected fluorescence is from NV$^0$. (This is smaller than what might be naively expected from Fig.~\ref{Fig1}b due to the possibility of recombination followed by NV$^-$ excitation within a single pulse, leading to collection of NV$^-$ fluorescence during the nominally ``NV$^0$" measurement.) The errors on all results come from statistical confidence intervals given by the fit, as well as considering variations in the literature values for NV lifetimes; since all rates are linear in green power, our results do not depend on our precise choice for pulse width.

%Lastly as an example, with the chosen pulse width of 100 ps we obtain a NV- excitation scaling factor of R = 223 ± 22 MHz/uW. This result will change with chosen pulse width while all the other results are independent of chosen pulse width (cut this?).

%(This corresponds to applying a square pulse of green illumination.)

\section{Stimulated emission and charge state switching}

We now move on to investigating the red-laser-induced effects of stimulated emission, ionization and recombination. Red illumination (700-800 nm) is known to cause depletion of NV fluorescence, but this reduction could come from either stimulated emission or charge state switching. % as both are non-radiative processes. 
Stimulated emission is the preferred mechanism as it is expected to conserve spin polarization in NV$^-$~\cite{wildanger_diffraction_2011}, while ionization does not~\cite{chen_spin_2015}; furthermore, stimulated emission underlies potential laser operation with NVs for which ionization would be a competing process. Earlier measurements on NV ensembles indicated minimal ionization~\cite{jeske_stimulated_2017}, but did not quantify the relative rates of stimulated emission and charge state switching. Our pulsed-laser approach allows us to corroborate these ensemble measurements and quantify the effects of red illumination on both NV$^-$ and NV$^0$ states.

The energy of the red light is insufficient to promote either charge state of the NV center to an excited state, so another source of excitation is required before the red laser effects can be investigated. We use a single pulse of green excitation to probabilistically populate the excited states, and subsequently apply a single red pulse. Since we already have data constraining the effects of the green excitation, we can isolate the transitions induced by the red pulse. These green-red pulse pairs were initially chosen to have a $592\pm2$ ps separation time, so that minimal excited-state decay will occur between pulses. 
  
\begin{figure}
	\includegraphics[scale=0.213]{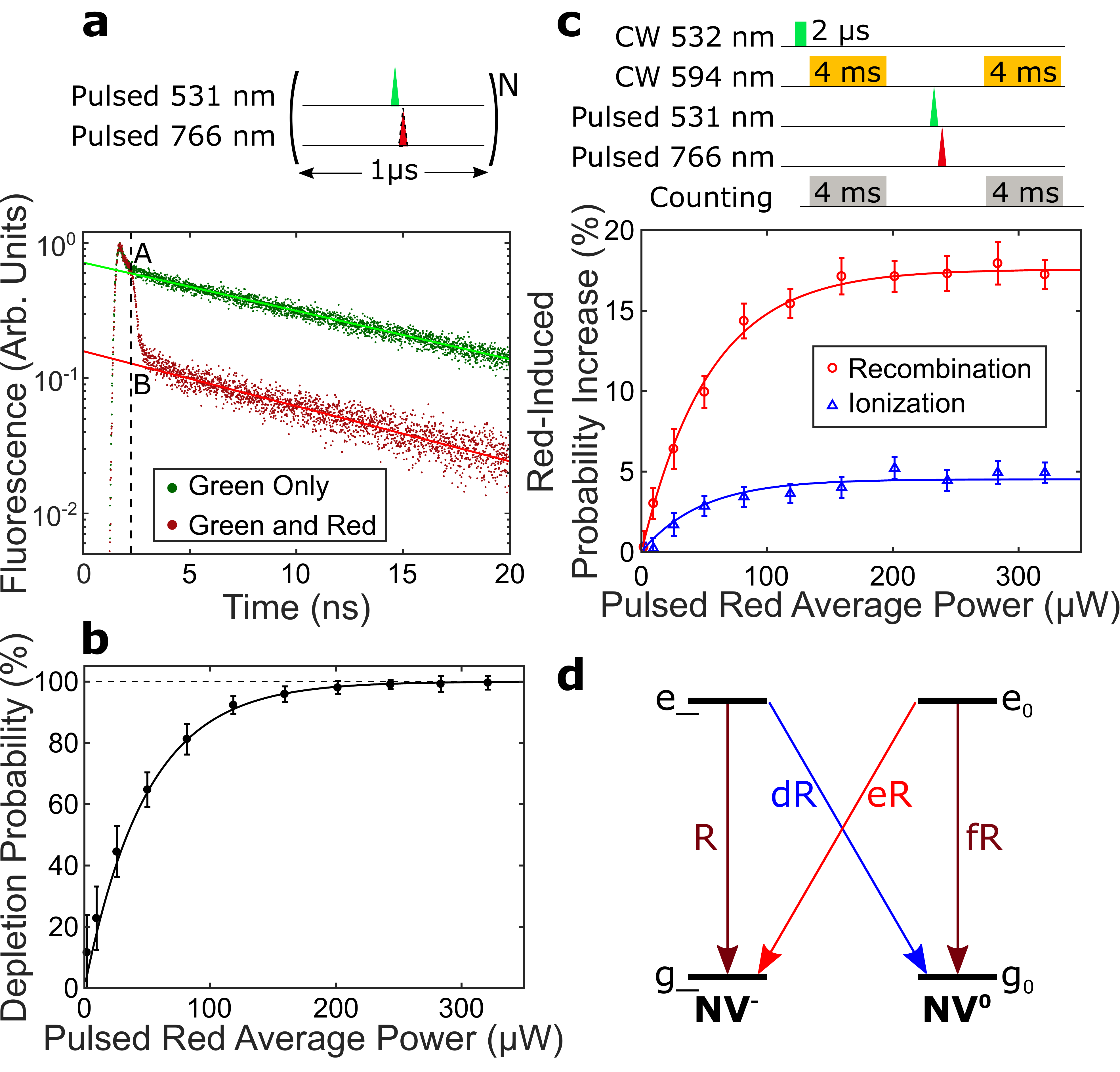}
	\caption{\label{Fig2} Stimulated emission and charge state switching under red illumination. \textbf{(a)} Top: Pulse sequence used for data shown in (a) (both with and without the red pulse)  and (b) (with both red and green pulses). The delay between the pulses is 0.592 ns. Bottom: Example fluorescence depletion data with pulsed green only (green, top) and green-red pulse pair (red, bottom) illumination. Solid lines are exponential decay fits. Extrapolating these fits to the location of the red pulse (dashed vertical line) gives the values marked as A and B. \textbf{(b)} NV excited state depletion probability (through stimulated emission or charge state switching) calculated by $1-\frac{B}{A}$ as a function of red power. \textbf{(c)} Top: Pulse sequence used for data shown at bottom; the green-red pulse delay is 0.592 ns. Bottom: Red-induced increase in the probability of ionization (blue triangles) and recombination (red circles),  measured by subtracting the charge state switching probability of a single green pulse from that of a green-red pulse pair. \textbf{(d)} Four-level rate equation model comprising the ground and excited states of NV$^-$ and NV$^0$. Here we show the transition rates under red illumination. Solid lines in \textbf{b} and \textbf{c} are found by using the four-level model to simultaneously fit data shown in Figs.~\ref{Fig1} and \ref{Fig2}.}
\end{figure}

We begin by investigating red-induced fluorescence-depletion as a function of red power. We measure the temporally-resolved NV fluorescence using a time-correlated single photon counting system (PicoHarp 300) synced to the pulsed laser output.  In Fig.~\ref{Fig2}a we show example time-resolved traces for pulsed green-only illumination (green, top, 95 $\upmu$W average green power), and green-red pulse pair illumination (red, bottom, 82 $\upmu$W average red power) with a 1 MHz repetition rate. The green-only data reveals the expected exponential decay of fluorescence, with a lifetime of $12.2\pm0.1$ ns, while the additional red pulse induces a sharp drop in fluorescence. After subtracting off background counts, we fit the long-time exponential decays for both the green-only and green-red data sets. These fits (solid lines in Fig.~\ref{Fig2}a) provide the fluorescence values that would occur at the location of the red pulse (Fig.~\ref{Fig2}a dashed vertical line), labeled A for green-only and B for green-red. %(Here we effectively neglect the duration of the red pulse in comparison to the spontaneous emission lifetime.) 
The proportional drop in fluorescence, given by $1-\frac{B}{A}$, gives a measure of the probability that a single red pulse causes the system to leave the NV excited states, either through charge state switching or stimulated emission, and will be referred to as the depletion probability. %We use error propagation on the exponential fit parameters to determine the error on the resulting depletion probability. <-- this is standard practice, doesn't need to be explained.
The depletion probability as a function of pulsed red power, shown in Fig.~\ref{Fig2}b, was found to saturate near 100$\%$ depletion with increasing red power. 

%(unable to fit with double/triple exponential to get NV0 decay…).

%(should say this is a continuous measurement, ie steady state…)

We now employ the pulse sequence shown in Fig.~\ref{Fig2}c to investigate whether red illumination causes ionization and recombination.  Working at the same fixed green and varying red powers used in the fluorescence depletion measurements, we observe that 766 nm illumination can cause increased charge state switching, above what was caused by the green excitation pulse; we quantify this by calculating the probability of ionization under green-red pulse pair illumination minus the probability of ionization with the green pulse alone. As shown in Fig.~\ref{Fig2}c, the red-induced probability increases in ionization (blue triangles) and recombination (red circles) saturate with increasing red power. This occurs because the red light also induces stimulated emission; once the NV is in the ground state, 766 nm illumination cannot cause further ionization/recombination.  %with recombination having a larger increase than ionization <-- there's really not much point in emphasizing this at this point, since it doesn't lead to any immediate conclusion.

The solid lines in Figs.~\ref{Fig2}b/c are simultaneous fits to the data of Figs.~\ref{Fig1} and \ref{Fig2} using the previously-described four-level model, with the relevant rates for red-induced stimulated emission ($R$ for NV$^-$, $fR$ for NV$^0$), ionization ($dR$)  and recombination ($eR$) shown in Fig.~\ref{Fig2}d. As for green, all rates are linear in red power $\propto R$. %As mentioned above, the red-dependent data (Figs.~\ref{Fig2}b/c) is fit simultaneously with the green-dependent data in Fig.~\ref{Fig1}. 
Simultaneous fitting requires that we correlate the green powers used in the two data sets, and we find that the green power used in the measurements of Figs.~\ref{Fig2}b/c corresponds to an average green power of 95 $\upmu$W as shown in Fig.~\ref{Fig1}~\footnote{Relating the green power used in Fig.~\ref{Fig2} to the green powers in Fig.~\ref{Fig1} is tricky because simple power-meter measurements may not accurately reflect the amount of power reaching the NV center due to drifts in alignment. Moreover, the fluorescence rate is relatively insensitive to green power at high power. Thus we use the green-induced ionization and recombination rates we record for the data of Fig.~\ref{Fig2} to identify the corresponding power in Fig.~\ref{Fig1}.}.
%The fixed green power used during the red-dependent experiments is calibrated by comparing its ionization and recombination probabilities to the previously measured charge state switching data in Fig.~\ref{Fig1}c. Such a calibration procedure is necessary because simple power-meter measurements may not accurately reflect the amount of power reaching the NV center due to drifts in alignment. With this calibration we can now model the effects of the initial green pulse on the NV center system from an arbitrary initial state. The next step is to use the data shown in Fig.~\ref{Fig2} to constrain the red-induced transition rates. 

We model the depletion probability data shown in Fig.~\ref{Fig2}b by calculating the fluorescence intensity immediately before and after the red pulse. We begin by initializing the model populations in the ground states of NV$^-$ and NV$^0$, with an appropriate charge state distribution found independently by applying a sequence of pulses with the same fixed green and varying red powers, and then reading out the charge state (data not shown). The model evolves under green illumination for 100 ps, probabilistically exciting both charge states. The excited states then relax for 592 ps, corresponding to the separation time between the green and red pulses. %Here we use the previously-mentioned NV$^-$ and NV$^0$ lifetimes ($12.2\pm0.1$ and $18\pm3$ ns); due to the short separation time between pulses in these experiments, varying the lifetime values does not significantly affect our results. %The excited state populations of NV$^-$ and NV$^0$ after this decay step are then added together, scaled by the same fluorescence scaling factors found in the green-dependent fits, to obtain a model value for the green-only fluorescence level at the time of the red pulse (labeled A in Fig.~\ref{Fig2}a). 
The green-only fluorescence level just before the red pulse (point A in Fig.~\ref{Fig2}a) is then given by $A = \alpha_- e_-(t_A) + \alpha_0 e_0(t_A)$, where the excited state populations $e_{-}(t)$ and $e_{0}(t)$ are evaluated at $t_A = $ 592 ps after the green pulse and $(\alpha_{-}, \alpha_0)$ are the same fluorescence scaling factors used to model the data of Fig.~\ref{Fig1}b.
Next, the model evolves for 100 ps under red illumination (with rates as in Fig.~\ref{Fig2}d), causing stimulated emission and charge state switching to occur. %The final excited state populations are again scaled and added together (using the same fit parameters as for Fig.~\ref{Fig1}b) to calculate the green-red fluorescence level at the time of the red pulse (labeled B in Fig.~\ref{Fig2}a). 
The fluorescence level immediately after the red pulse (point B in Fig.~\ref{Fig2}a) is $B = \alpha_- e_-(t_B) + \alpha_0 e_0(t_B)$, now with the populations $e_{-,0}(t)$ evaluated at $t_B$, immediately after the red pulse. 
The resulting value $1-\frac{B}{A}$ %is calculated with varying red power ($\propto R$) in Fig.~\ref{Fig2}d) and
 is used to fit the measured depletion probability.

The red-induced increase in the charge state switching probabilities shown in Fig.~\ref{Fig2}c is fit in a similar manner. The model is initialized entirely into one charge state (in the ground state), evolved under the calibrated green pulse, a 592 ps decay time, and a final red pulse, and the resulting population in the opposite charge state (ground plus excited states) is extracted. The simultaneous fit provides the red-dependent rates, which we again quote as a percentage of a given rate, here the NV$^-$ stimulated emission rate ($R$ in Fig.~\ref{Fig2}d). We obtain relative rates of red-induced ionization $d=7.1\pm0.3\%$, recombination $e=22\pm2\%$ and NV$^0$ stimulated emission $f=74\pm6\%$. These results indicate that our red pulses are mainly -- but not entirely -- inducing stimulated emission. 

%Lastly as an example, with the chosen pulse width of 100 ps we obtain a NV- stimulated emission scaling factor of R = 192 ± 8 MHz/uW (cut?). 

%Also considered ionization to excited state of NV0… Changes red-induced ionization fit to 9.7 pm 4$\%$ of stimulated emission.

\section{Green power dependence of red behaviour}

Recent literature results have shown that the effects of red (or near-IR) illumination on the NV center charge state distribution strongly depend on green excitation power \cite{hopper_near-infrared-assisted_2016,ji_charge_2016}.  In particular, a red-induced increase of NV$^-$ population occurs when exciting with low green power, while an induced decrease occurs at high green powers. 
  
  \begin{figure}
	\includegraphics[scale=0.22]{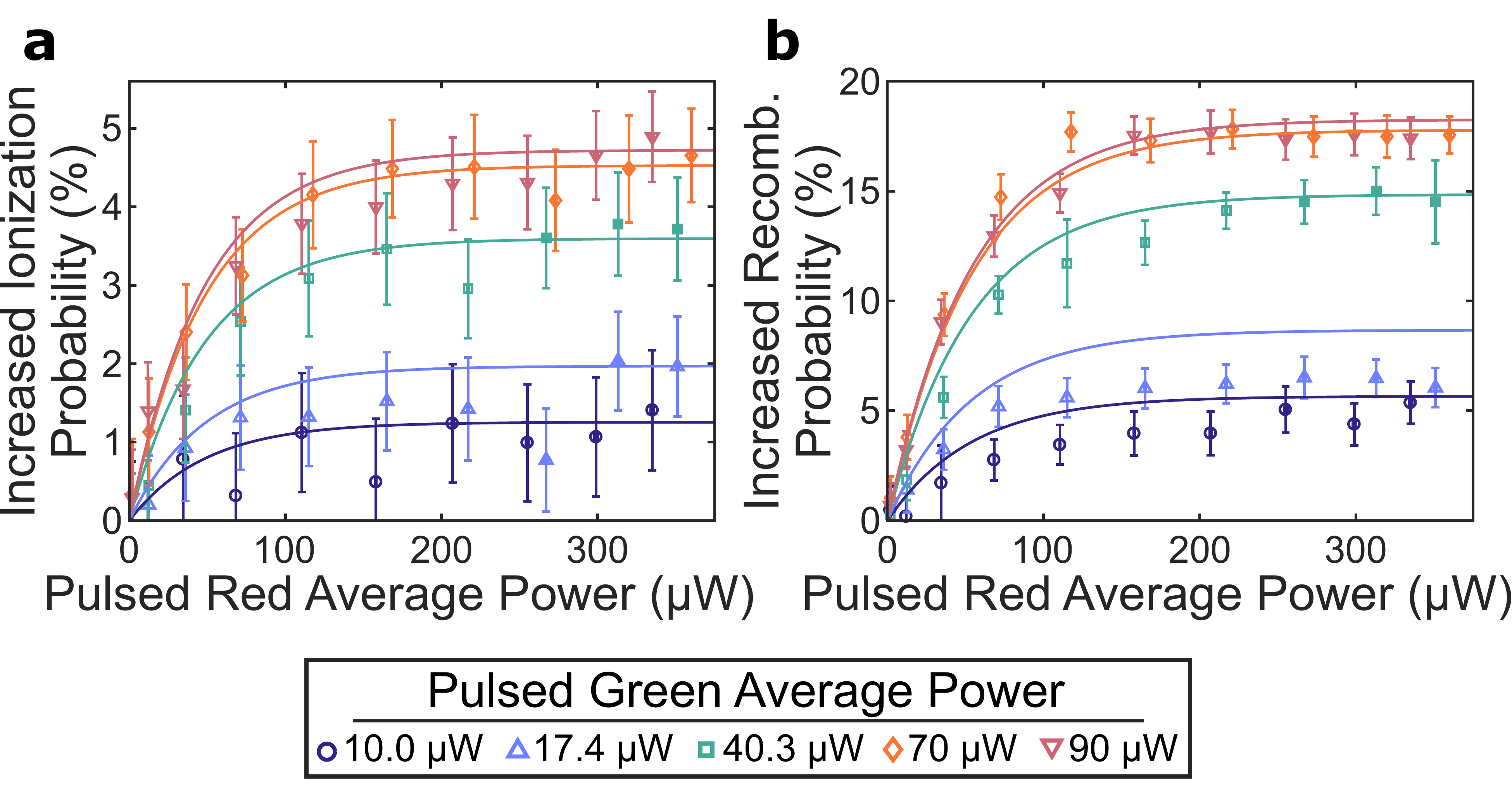}
	\caption{\label{Fig3} Green and red power dependence on charge state switching rates. Red-induced increase in the probability of ionization (\textbf{a}) and recombination (\textbf{b}) are plotted as a function of red power for the five green powers shown in the legend. Solid lines are model predictions at the same powers, using rates determined from fits to figures \ref{Fig1} and \ref{Fig2}. All powers are given at 1 MHz repetition rate.}
\end{figure}

  To examine if any of these non-monotonic changes occur in the single pulse regime, we probe the charge state switching probabilities of green-then-red pulse pairs (using the same pulse sequence as in Fig.~\ref{Fig2}c), with varying green and red power. Figure 3 shows the red-induced increase in charge state switching probabilities (above the green-only probability, which follows Fig.~\ref{Fig1}c). %The results are shown in Fig.~\ref{Fig3}, in which we plot the red-induced increase in charge state switching probabilities compared to the green-only probabilities, which follow Fig.~\ref{Fig1}c. 
  We find that the red-induced increase in ionization and recombination probabilities monotonically increases with green power.  Physically, this makes sense: increased green power results in more excited state population that the red illumination can affect. %This is caused by the increased excitation rate with green power, resulting in more excited state population that the red illumination can affect. We also observe that the charge state switching probabilities saturate with increasing red power regardless of the green power. 
However, this monotonic behavior would predict a saturating NV charge state distribution, which is not observed in the literature. 

The resolution lies in the fact that we are working with single pulse pairs. If we examine steady state behavior instead, our results agree with previous measurements (see appendix~\ref{AppendixB}). Thus, while the underlying ionization behavior is quite simple (as measured in our single-pulse experiments), longer time-scale dynamics, such as those associated with the NV$^-$ singlet states and spin states, play a critical role in determining nontrivial steady-state charge distributions. These results thus highlight the utility of single-pulse experiments in isolating intrinsic optical excitation effects from slowly-evolving internal dynamics.

The solid lines in Fig.~\ref{Fig3} show predictions of our four-level model, using rates extracted from fits to Figs.~\ref{Fig1} and \ref{Fig2}. These predictions are in reasonable agreement with our data, and discrepancies are likely due to drifts in the alignment of the green laser during data acquisition over a period of a week. The model predicts that at even higher green powers, the maximum red-induced increase in the charge state switching probabilities will decrease, as green-induced ionization and recombination reduces the population in excited states from which the red pulse can affect the system.

\begin{figure*}
	\includegraphics[scale=0.23]{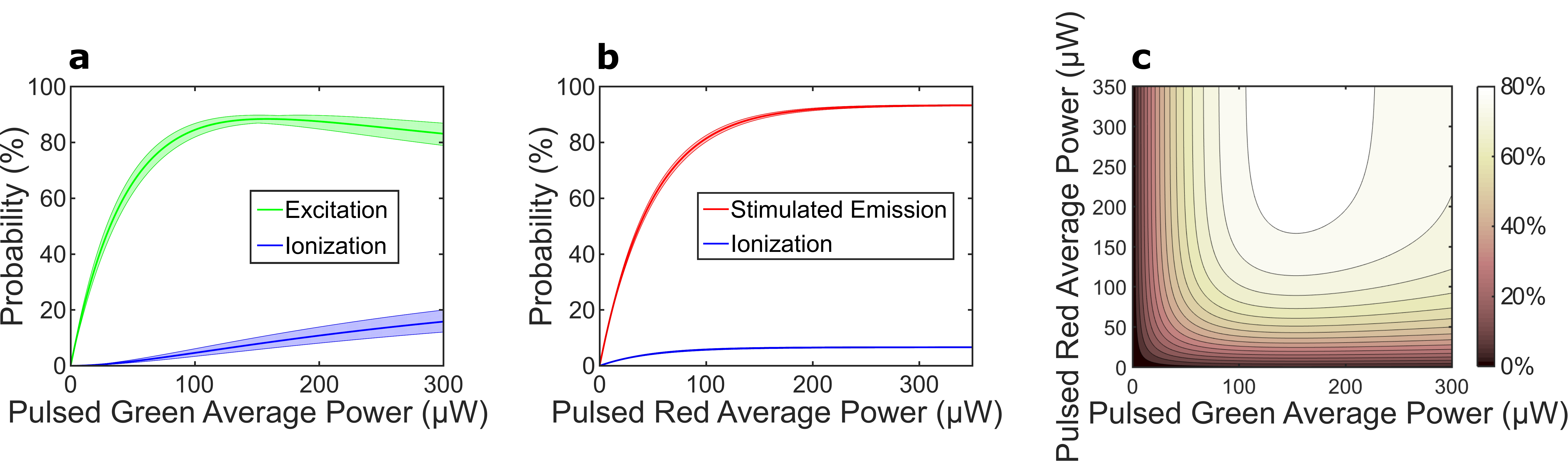}
	\caption{\label{Fig4} Four-level model predictions. \textbf{(a)} Starting in the NV$^-$ ground state, probability of green-induced excitation (green top) versus ionization (blue bottom). \textbf{(b)} Starting in the NV$^-$ excited state, probability of red-induced stimulated emission (red top) versus ionization (blue bottom). \textbf{(c)} Starting in the NV$^-$ ground state, cycling probability of excitation followed by stimulated emission. Shaded regions in \textbf{a} and \textbf{b} represent one standard error.}
\end{figure*}

\section{Model Predictions}

We now use the four-level model fit results to predict the effects of single green pulses on the NV$^-$ ground state, and single red pulses on the NV$^-$ excited state.  Assuming that the system starts entirely in the NV$^-$ ground state, Fig.~\ref{Fig4}a shows the predicted probability to end up in the NV$^-$ excited state (green, top) versus ionizing to NV$^0$ (blue, bottom) after a single green pulse (shaded region shows one standard error). The excited state probability initially grows due to an increased excitation rate, but eventually starts to decrease as ionization takes over. This leads to an optimum power for maximizing excited state population after a single green pulse, which was found to be 158 $\upmu$W average pulsed power (at 1 MHz repetition rate), leading to an excited state probability of $88.4\pm1.5\%$.

%??? Should we instead estimate the energy per pulse? 

Starting entirely in the NV$^-$ excited state, Fig.~\ref{Fig4}b shows the predicted probability to end up in the NV$^-$ ground state via stimulated emission (red, top) versus ionizing to NV$^0$ (blue, bottom) after a single red pulse. Both stimulated emission and ionization saturate with increasing red power. At saturating powers (350 $\upmu$W), there is a predicted $93.3\pm0.2\%$ probability of stimulated emission, and a $6.6\pm0.2\%$ probability of ionization.

Excitation followed by stimulated emission has potential applications as a proposed mechanism to engineer fast spin gates between electron and nuclear spins~\cite{fuchs_excited-state_2010}, and could also be useful in detecting joint states of closely-spaced spins using stimulated emission depletion (STED) microscopy~\cite{rittweger_sted_2009}. %to resolve closely-spaced spins, %proposed laser threshold magnetometry \cite{Jeske2016} (cut?), 
Such applications ideally seek to maintain spin coherence or polarization during the excitation and stimulated emission processes, which can be lost through ionization~\cite{chen_spin_2015}. In Fig.~\ref{Fig4}c we examine the predicted probability to fully cycle within NV$^-$ (ground to excited state and back) for varying pulsed green and red powers. We find that a 153 $\upmu$W pulsed green average power and a maximum pulsed red average power (350 $\upmu$W) results in an optimum cycling probability of $82.5\pm0.3\%$. As another example, our model predicts that with an initial $80\%$ NV$^-$ charge state population, and $90\%$ spin polarization, an optimized green-then-red pulse sequence will result in a final $78\pm5\%$ NV$^-$ population with $81\pm9\%$ spin polarization, in rough agreement with the preservation of spin polarization during STED found by Wildanger {\it et al.}~\cite{wildanger_diffraction_2011}.

\begin{table*}[ht]
	\centering
	\caption{\label{Table1} Summary of transition rates}
	\begin{tabular}{ l | c | c }
		\hline\hline
		& Pulsed 531 nm  & Pulsed 766 nm  \\
		& ($\%$ of $NV^-$ excitation rate $G$) & ($\%$ of $NV^-$ stimulated emission rate $R$) \\
		\hline
		Ionization & $a=3.7\pm0.6\%$ & $d=7.1\pm0.3\%$ \\
		%\hline
		Recombination & $b=8\pm1\%$ & $e=22\pm2\%$ \\
		%\hline
		$NV^0$ excitation & $c=130\pm20\%$ & \\
		%\hline
		$NV^0$ stimulated emission & & $f=74\pm6\%$ \\
		\hline
		$NV^-$ singlet ionization & & $I_s/R = 2.15\pm0.5 \%$  at $R = R_0=66.9\pm0.3$ GHz\\
		\hline \hline
		%\cline{1-3}\hhline{==}
		%&\multicolumn{1}{|c}{Lifetimes (ns)}& \\
		%\cline{1-2}
		%$NV^0$ & \multicolumn{1}{|c}{$\frac{1}{\eta}=15.9\pm1.5$} & \\
		%$NV^-$ Singlet & \multicolumn{1}{|c}{$\frac{1}{D}=141\pm20$} & \\
		%\cline{1-2}
	\end{tabular}
	\label{tab:AOMs}
\end{table*}

\section{Varying pulse separation}

The short pulse separation time used during the above experiments ensured that we were primarily probing charge state switching out of the NV$^-$ and NV$^0$ excited states. By increasing the separation time between the green and red pulses we can investigate ionization and recombination out of metastable states that exist in either charge state.  Red-induced ionization out of the metastable singlet state in NV$^-$ has recently been of interest for use in spin-to-charge conversion \cite{hopper_near-infrared-assisted_2016}; our technique could also reveal recombination from metastable states in NV$^0$, such as the predicted quartet state \cite{felton_electron_2008}. 

\begin{figure}
	\includegraphics[scale=0.213]{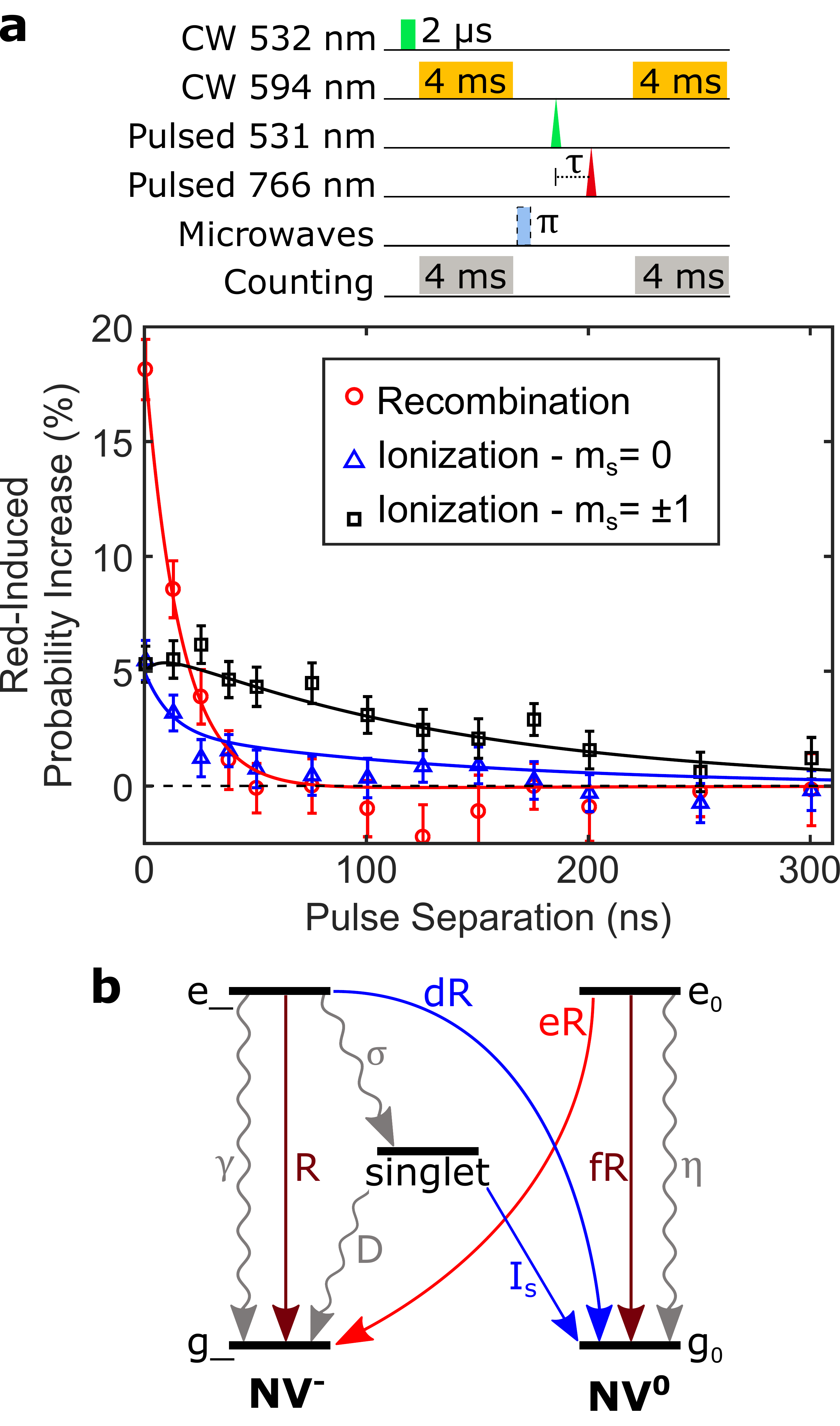}
	\caption{\label{Fig5} Red-induced charge state switching versus pulse separation. \textbf{(a)} Top: Pulse sequence used, with a variable delay $\tau$ between green and red pulses. A microwave $\pi$ pulse is optionally applied to initialize into the $m_s = \pm1$ states. Bottom: Red-induced increase in the probability of recombination (red circles) and spin-dependent ionization for a $m_s=0$ (blue triangles) and $m_s=\pm1$ (black squares) initial NV$^-$ spin state. \textbf{(b)} Five-level rate equation model consisting of the ground and excited states of NV$^-$ and NV$^0$ as well as a level for the NV$^-$ singlet state. The squiggly  arrows indicate relaxation rates that are relevant during the separation time, including shelving ($\sigma$), deshelving ($D$), and spontaneous emission $\gamma$ ($\eta$) from NV$^-$ (NV$^0$); straight and curved lines are  red-induced transitions, with an additional ionization rate $I_s$ out of the NV$^-$ singlet state. Note that the value of $\sigma$ is different for the two spin states.}
\end{figure}

We use the pulse sequence shown in Fig.~\ref{Fig5}a (top) to measure the red-induced increase in the ionization and recombination probabilities as a function of pulse separation time $\tau$. Due to the maximum 80 MHz repetition rate of the pulsed laser driver internal oscillator, the minimum step-size for pulse separation is 12.5 ns. 
In order to fully explore the NV$^-$ singlet ionization dynamics, we use microwave $\pi$ pulses to flip the NV$^-$ spin, which is initially mostly polarized into $m_s=0$, into the $m_s=\pm1$ spin states. The $m_s=\pm1$ spins have a much higher probability of decaying through the NV$^-$ singlet state, allowing us to observe the role of the singlet state in the ionization dynamics.

The results of this experiment are shown in Fig.~\ref{Fig5}a (bottom). The recombination probability out of NV$^0$ decreases exponentially with approximately the excited state lifetime, indicating that there are no long-lived NV$^0$ metastable states from which our red laser can induce significant recombination. There is, however, a significant difference in the $\tau$-dependence of the red-induced ionization probabilities for the two NV$^-$ spin states. For $m_s=0$, ionization decreases exponentially with the NV$^-$ excited state lifetime, while the for $m_s=\pm1$ a much longer decay time is observed, associated with the metastable singlet state.

We fit our results using a five-level rate equation model as shown in Fig.~\ref{Fig5}b, now with the inclusion of a level associated with the NV$^-$ singlet state. Similar to the previously described fitting procedures, this data is fit simultaneously with the data shown in Figs.~\ref{Fig1} and \ref{Fig2}, and the green and red powers are correlated across data sets using ionization and recombination rates. The inclusion of the NV$^-$ singlet state and spin-dependence requires that some parameters be fixed to literature values to reduce our parameter space. We fix the NV$^-$ $m_s=\pm1$ excited state lifetime to $6.0\pm0.1$ ns (using the ratio of $m_s = 0$ to $m_s = \pm 1$ lifetimes from~\cite{gupta_efficient_2016}); we set the excited state to singlet state decay probability for $m_s=0$ spins to $15\pm5\%$ and for $m_s=\pm1$ spins to $55\pm3\%$ \cite{gupta_efficient_2016,robledo_spin_2011}; together with the measured $m_s = 0$ excited state lifetime, these parameters determine the shelving rates $\sigma$ for the two spin states. Lastly, we fix the spin polarization to $90\pm10\%$~\cite{doherty_nitrogen-vacancy_2013}. The errors on these fixed parameters are considered during error analysis procedures.

We model our pulse separation data by initializing entirely into the ground state of either charge state, applying a 100 ps green pulse (using the four-level model in Fig.~\ref{Fig1}d), letting the excited states spontaneously decay for a duration equal to the pulse separation (and allowing decay into and out of the NV$^-$ singlet state), and then applying a 100 ps red pulse (relevant rates shown in Fig.~\ref{Fig5}b). The fit results for the green and red parameters describing excitation, stimulated emission, and ionization/recombination out of the excited states all lie within error of the previously quoted results (see Table~\ref{Table1}). 

The new fit results consist of the NV$^0$ excited state lifetime of $1/\eta = 15.9\pm1.5$ ns, the NV$^-$ singlet lifetime of $1/D =141\pm20$ ns, and the red-induced ionization rate $I_s$ out of the singlet state.   Because we only measure singlet ionization at a single power, our data does not constrain the functional dependence of $I_s$ on $R$ (which likely varies from quadratic to linear as $R$ increases, due to saturable absorption within the singlet manifold~\cite{kehayias_infrared_2013}). Nevertheless, we can compare $I_s$ to the NV$^-$ stimulated emission rate at this specific power, where $R = R_0 = 66.9\pm 0.3$ GHz; we find that $I_s$ is $2.15\pm0.5\%$ of $R_0$. Comparing to the previously quoted red-induced ionization rate out of the excited state of $7.1\pm0.3\%$ of $R$, one might expect that there would be more ionization occurring out of the excited state when compared to the singlet state. In fact this is not the case, as our model predicts that our red pulse induces an overall $6.6\pm 0.2\%$ probability of ionization if the system starts in the NV$^-$  excited state and a $14\pm3\%$ probability of ionization out of the singlet state. The reason for this apparent discrepancy is that ionization out of the excited state has to compete with stimulated emission, while singlet ionization does not.

\section{Conclusion and Outlook}

Our experiments add to a growing body of work to understand excitation, depletion, ionization, and recombination of the NV center under green and red (or near-IR) illumination. Notably, by working with single green and red pulses, with control over the initial charge and spin states, we are able to greatly simplify the modeling and extract quantitative rates for optically-induced charge-state switching out of the singlet and excited states. These models are simple to use and provide a basis for understanding more complex behaviours. For example, our model places constraints on proposed fast electron-nuclear spin quantum gates using enhanced excited-state hyperfine interactions~\cite{fuchs_excited-state_2010}, and indicates limits to the spin-polarization-preserving properties of STED~\cite{wildanger_diffraction_2011, chen_spin_2015}. We also make a quantitative measurement of the relative rates of stimulated emission and ionization induced by red illumination of NV$^-$, which corroborates earlier ensemble measurements~\cite{jeske_stimulated_2017} and provides a parameter for calculating lasing thresholds~\cite{jeske_laser_2016}. Moreover, our measurement of relative ionization rates out of NV$^-$ excited and singlet states is potentially relevant to spin-to-charge conversion~\cite{hopper_near-infrared-assisted_2016}. Ultimately, an improved understanding of the fundamental optically-induced dynamics of the NV center can impact a broad range of current and future applications.

%Conclusion
%- Extracted rates that allow for simple modeling of ionization/recombination under green and red illumination
%- Provides ionization-based constraints on proposed electron-nuclear spin gates
%- Also for STED-based readout of nearby spins
%- corroborates that stimulated emission dominates (but ionization isn't zero) for laser threshold magnetometry
%- On longer timescales reveals relative ionization rates for red out of excited vs singlet states -- role played by stimulated emission potentially relevant to spin-to-charge conversion

%Outlook
%??

\begin{acknowledgments}
LC acknowledges funding support from Canada Foundation for Innovation and Canada Research Chairs project 229003, Fonds de Recherche - Nature et Technologies FQRNT NC-172321, National Sciences and Engineering Research Council of Canada NSERC RGPIN 435554-13, and l'Institut Transdisciplinaire d'Information Quantique (INTRIQ). 
\end{acknowledgments}

\appendix
\section{Extracting ionization and recombination probabilities}
\label{AppendixA}

\renewcommand{\thefigure}{A\arabic{figure}}
\setcounter{figure}{0}

High fidelity charge state initialization and readout is performed using yellow (594 nm) illumination. We model the resulting photon count distribution using the techniques described by Shields {\it et al.} \cite{shields_efficient_2015}. This approach assumes that the dynamics can be fully described by the rates of ionization, recombination and the fluorescence rates for the two charge states, considering all possible charge state switching sequences and weighting them by their probability to occur. For a given initial charge state that undergoes switching and spends a time $\tau$ in the initial state and $t_R - \tau$ in the other charge state, the photon count distribution is given by the convolution of two Poisson distributions, one corresponding to photons emitted during time $\tau$ in the initial state, and the other corresponding to photons emitted during time $t_R - \tau$ in the other charge state. The counts from each charge state follow a standard Poisson distribution, which we write as $Poiss(n,x)$, representing the probability of obtaining $n$ counts given the average number of counts $x$. 
With a fixed measurement time $t_R$, ionization rate $\Gamma_-$, recombination rate $\Gamma_0$, NV$^-$ fluorescence rate $\mu_-$, and NV$^0$ fluorescence rate $\mu_0$, we calculate the following photon count distribution given that we are initially in the NV$^-$ charge state: $P(n|NV^-) = P(n|NV^-, even) + P(n|NV^-, odd)$, where

\begin{widetext}
	\begin{equation}	
	\label{shields}
	\begin{aligned}
	%P(n\lvert NV^-,0) &= e^{-\Gamma_-t_R}Poiss(n,\mu_-t_R)\\
	P(n\lvert NV^-,even) = &\int_{0}^{t_R}d\tau e^{-\Gamma_-\tau-\Gamma_0(t_R-\tau)}Poiss(n,\mu_-\tau+\mu_0(t_R-\tau))\sum_{m=1}^{\infty}\frac{\Gamma_0^m\Gamma_-^m\tau^m(t_R-\tau)^{m-1}m}{m!^2}\\
	&+e^{-\Gamma_-t_R}Poiss(n,\mu_-t_R)\\
	P(n\lvert NV^-,odd) = &\int_{0}^{t_R}d\tau e^{-\Gamma_-\tau-\Gamma_0(t_R-\tau)}Poiss(n,\mu_-\tau+\mu_0(t_R-\tau))\sum_{m=1}^{\infty}\frac{\Gamma_0^{m-1}\Gamma_-^m\tau^{m-1}(t_R-\tau)^{m-1}}{(m-1)!^2},
	\end{aligned}
	\end{equation}
\end{widetext}

\noindent which has been written out for all even or all odd numbers of switching events (where $m$ represents the number of even or odd switching events). The last term in the expression for $P(n\lvert NV^-,even)$ is for zero switching events. These results can be rewritten using modified Bessel functions of the first kind to obtain the exact results quoted by Shields {\it et al.}~\cite{shields_efficient_2015}. The distribution for initialization in NV$^0$ is found by simply switching $-\leftrightarrow0$ in the above equations. The final expected distribution is found by adding the distributions for an initial state of NV$^0$ and NV$^-$, scaled by the probability to start in each charge state: 
\begin{equation}
P(n) = P(n|NV^-)*P(NV^-) + P(n|NV^0)*P(NV^0),
\end{equation}
where $P(NV^-)$ and $P(NV^0)$ are the probabilities that the NV starts in the NV$^-$ or NV$^0$ charge state at the beginning of the yellow pulse.

\begin{figure}
	\includegraphics[scale=0.22]{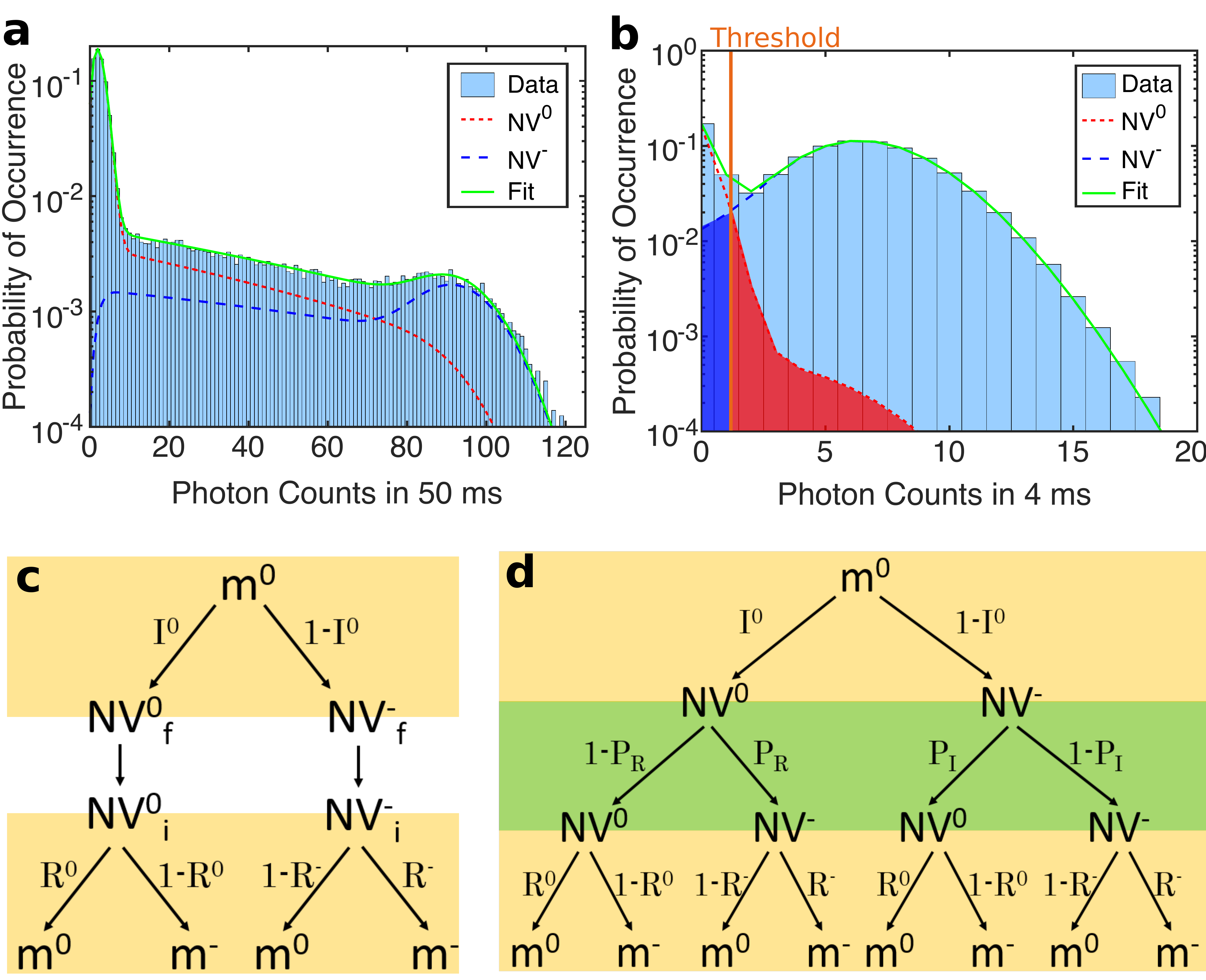}
	\caption{\label{A1} Extracting ionization and recombination probabilities. \textbf{(a)} Photon count distribution from 1 hour of 2 $\upmu$W continuous yellow illumination in 50 ms bins. Fitting (green solid line) allows for extraction of the yellow charge state switching rates. \textbf{(b)} Typical photon count distribution of a 4 ms, 2 $\upmu$W yellow pulse after initializing the charge state distribution with a pulse of CW green. Fits allow us to extract the yellow readout fidelities. \textbf{(c)} Probability tree representing processes with no applied experiment between two yellow pulses. Expressions next to arrows indicate the probability of a given process, e.g. $I^0$ is the probability to end up in state NV$^0_f$ given an initial measurement of $m^0$. \textbf{(d)} Probability tree representing measurements from two yellow pulses with an applied experiment (green row), allowing for extraction of the probability to ionize ($P_I$) and recombine ($P_R$).  For both (c) and (d), a similar tree can be constructed for an initial measurement $m^-$ by exchanging $0 \leftrightarrow -$.}
\end{figure}

Using this model we can extract the charge state switching rates $\Gamma_-$ and $\Gamma_0$ under continuous yellow illumination. This is done by collecting photon counts into 100~$\upmu$s bins for one hour, with further binning performed during post-processing. An example photon count distribution for one hour of 2 $\upmu$W yellow illumination with 50 ms binning is shown in figure \ref{A1}a.  We fit the data to Eq.~\ref{shields} to extract the charge state switching and photon emission rates, $\Gamma_0=3.89\pm0.07$ Hz, $\Gamma_-=27\pm1$ Hz, $\mu_0=48.87\pm0.09$ Hz, and $\mu_-=1870\pm30$ Hz. Here, since we are in a steady state situation, the probability to be in NV$^-$ is given by $\frac{\Gamma_0}{\Gamma_0+\Gamma_-} = 12.6\%$. When doing these fits we ensure that the chosen bintime is longer than $\frac{1}{\Gamma_-}$, allowing for sufficient switching events to occur within the bintime to give a good estimate for $\Gamma_-$. We use this procedure primarily to characterize the yellow charge state switching rates for later calculations.  

The actual pulses used for charge initialization and readout are shorter, and reflect a different initial charge state distribution. In figure \ref{A1}b we show a typical photon count distribution for a 4 ms, 2 $\upmu$W yellow pulse (duration and power chosen to maximize fidelities while minimizing duration) after initializing the charge state distribution with a pulse of CW green (i.e. from the first yellow pulse shown in figure \ref{Fig1}a). This distribution is fit using the same model discussed earlier but with fixed charge state switching rates (extracted as described above), and varying charge state distribution. For the example shown we obtain a $80.6\pm0.2\%$ NV$^-$ population. We now define a threshold, such that any counts less than or equal to the threshold gives result $m^0$ and any counts above gives result $m^-$. Using this threshold we want to determine the probability of correctly predicting the NV charge state at the beginning (readout) or end (initialization) of the yellow pulse.

The error in charge state readout stems from the fact that there is some overlap between the two charge state distributions, shown as shaded regions in figure \ref{A1}b. %We define the blue shaded region as $N^-\epsilon^-$ and the red shaded region as $(1-N^-)\epsilon^0$, where $N^-$ is the NV$^-$ population, and we can extract both $N^-$ and $\epsilon^{0, -}$ from the fit. 
We define the blue shaded region (below the threshold and under the blue dashed curve) as $N^-\epsilon^-$ and the red shaded region (above the threshold and under the red dotted curve) as $(1-N^-)\epsilon^0$, where $N^-$ is the NV$^-$ population, and we can extract both $N^-$ and $\epsilon^{0, -}$ from the fit. Our readout fidelity, depending on charge state, is then given by:

\begin{equation}
R^{0(-)}=P(m^{0(-)}\lvert NV^{0(-)}_i)=1-\epsilon^{0(-)}
\end{equation} 

\noindent which is the probability to obtain the correct result $m^{0(-)}$ given an initial $NV^{0(-)}$ state. Since the initial state is a given, our extracted readout fidelities will not change with shifts in charge state distribution, and will remain valid for the second yellow pulse in the sequence (fig. \ref{Fig1}a), regardless of experiment. We choose our threshold to maximize population-weighted average fidelity $1-\frac{(1-N^-)\epsilon^0+N^-\epsilon^-}{2}$.

In order to characterize our initialization fidelities, we measure the probability that two yellow pulses give the same result, with no applied experiment in between. %This consists of repeating the pulse sequence shown in figure \ref{Fig1}a with no pulsed illumination, and measuring if the counts are above or below the previously defined threshold. The non-demolition probability is then defined as 
We define $Q^{0(-)}=P(m^{0(-)}_2\lvert m^{0(-)}_1)$, which is the probability of obtaining the result $m^{0(-)}$ during the second yellow pulse given that we had the result $m^{0(-)}$ during the first yellow pulse.

We can expand $Q^{0(-)}$ in terms of initialization and readout fidelities, as shown in a probability tree in Fig.~\ref{A1}c. We define the initialization fidelities as $I^{0(-)}=P(NV^{0(-)}_f\lvert m^{0(-)})$, which is the probability of ending in the final charge state NV$^{0(-)}_f$ given that we measure $m^{0(-)}$. The final charge state of the first yellow pulse will be equal to the initial charge state of the second pulse due to slow charge state relaxation in the dark (measured to be on the order of a second). Reading off the possible paths to get from result $m^0$ in the first pulse to result $m^0$ in the second pulse gives:

\begin{equation}
Q^{0(-)}_Z = I^{0(-)}R^{0(-)}+(1-I^{0(-)})(1-R^{-(0)})
\end{equation}

\noindent where $Q^-_Z$ was found using a similar process, and the Z indicates no applied experiment between pulses. Solving for the initialization fidelities, we find:

\begin{equation}
I^{0(-)} = \frac{Q^{0(-)}_Z+R^{-(0)}-1}{R^0+R^--1}.
\end{equation}

We can now perform an experiment (e.g. apply pulsed illumination) between the two yellow pulses and use the characterized fidelities to calculate the probability that the experiment causes the NV center to switch charge states. Again, we compare two successive charge state measurements; now there is some probability of switching charge states between the two yellow pulses due to the applied experiment, which we define as $P_I=P(NV^0\lvert NV^-)$ for ionization and $P_R=P(NV^-\lvert NV^0)$ for recombination. An example probability tree for finding $Q^0$ is shown in Fig.~\ref{A1}d. Reading off the possible paths (with a similar method for finding $Q^-$) gives us:

%\begin{widetext}
	%\begin{equation}	
	%\begin{aligned}
	%Q^0 = I^0(1-P_R)R^0 + I^0P_R(1-R^-) + (1-I^0)P_IR^0 + (1-I^0)(1-P_I)(1-R^-)\\
	%Q^- = I^-(1-P_I)R^- + I^-P_I(1-R^0) + (1-I^-)P_RR^- + (1-I^-)(1-P_R)(1-R^0),
	%\end{aligned}
	%\end{equation}
	
	\begin{equation}
	\begin{aligned}	
	Q^0 =& I^0(1-P_R)R^0 + I^0P_R(1-R^-) \\ & + (1-I^0)P_IR^0 + (1-I^0)(1-P_I)(1-R^-)\\
	Q^- =& I^-(1-P_I)R^- + I^-P_I(1-R^0)\\& + (1-I^-)P_RR^- + (1-I^-)(1-P_R)(1-R^0),
	\end{aligned}
	\end{equation}
	
	\noindent which can be solved for $P_I$ and $P_R$:
	%\end{widetext}
	\begin{equation}
	\begin{aligned}
	P_I = \frac{I^0(Q^--R^-)+(1-I^-)(R^-+Q^0-1)}{(1-I^0-I^-)(1-R^0-R^-)}\\
	P_R = \frac{I^-(Q^0-R^0)+(1-I^0)(R^0+Q^--1)}{(1-I^0-I^-)(1-R^0-R^-)}.
	\end{aligned}
	\end{equation}

Since we have determined our initialization and readout fidelities $I^{0,-}$ and $R^{0, -}$, and we directly measure $Q^{0,-}$ from experiments using the pulse sequence shown in e.g. Fig.~\ref{Fig1}a, we can extract the ionization and recombination probabilities for an arbitrary experiment.

In summary our process of extracting charge state switching probabilities consists of: (a) characterizing the yellow charge state switching rates by fitting the photon count distribution from an hour of continuous yellow illumination data, (b) determining the yellow readout fidelities by fitting the photon count distribution of the first yellow pulse (in the pulse sequences shown in Figs. \ref{Fig1}a, \ref{Fig2}c, and \ref{Fig5}a), (c) finding the yellow initialization fidelities by measuring the probability that the two yellow pulses give the same result with no applied experiment, and finally (d) extracting the probability of ionization and recombination from charge state measurements before and after the applied experiment. We use standard error and error propagation to obtain the final errors on these values.

\section{Steady State Measurements}
\label{AppendixB}

\begin{figure}[h]
	\includegraphics[scale=0.205]{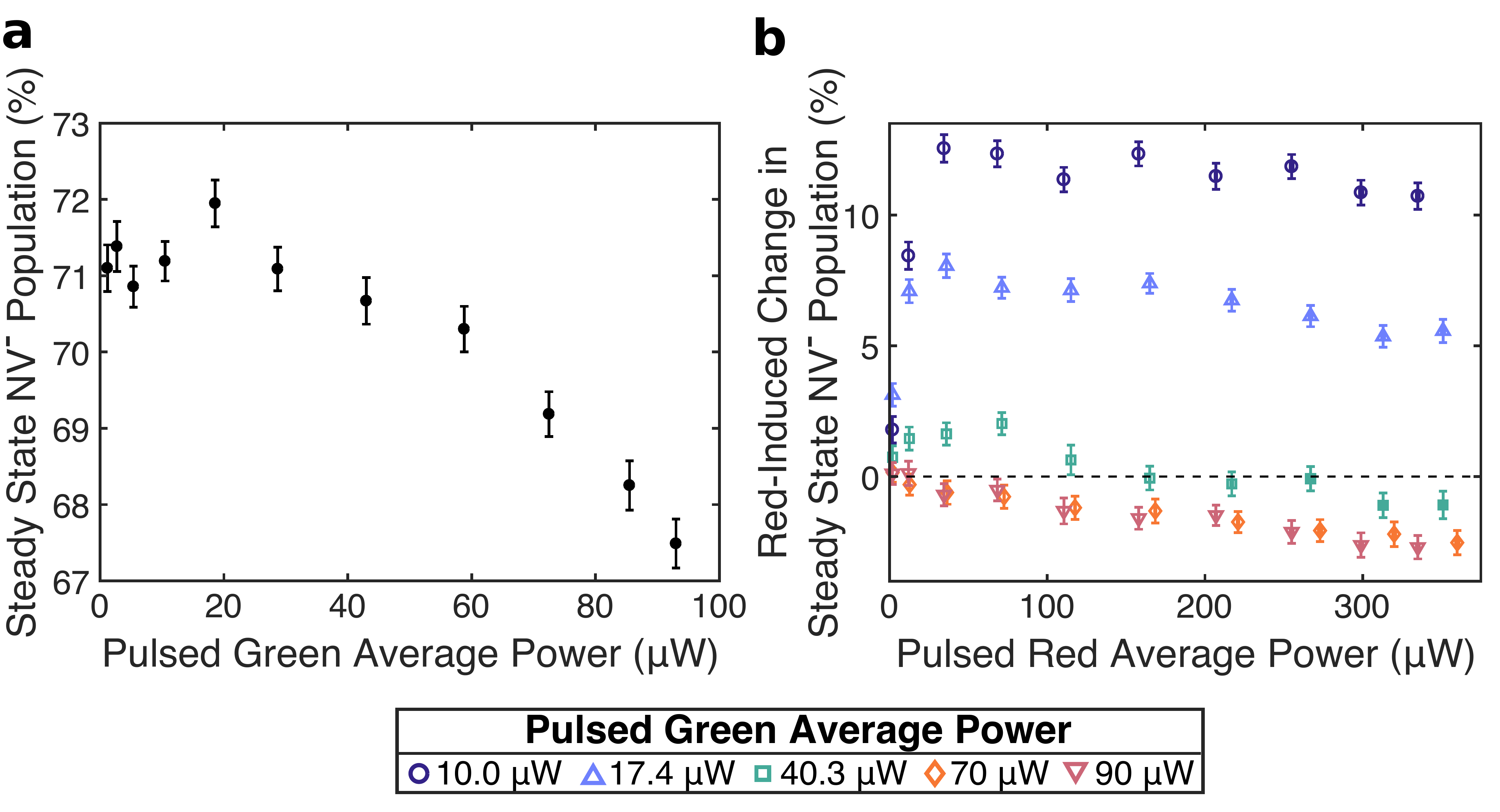}
	\caption{\label{A2} Steady state measurements. \textbf{(a)} Steady state NV$^-$ charge state population under pulsed green-only illumination, as a function of power. \textbf{(b)} Red-induced change in steady state NV$^-$ charge state population during pulsed green-then-red illumination, as a function of green and red power. Repetition rates are 1 MHz.}
\end{figure}

In contrast to our single pulse or pulse-pair experiments, we also measure steady state values after a long duration of pulsed illumination. Here we measure the final NV$^-$ charge state population after 150 $\upmu$s of pulsed illumination (1 MHz repetition rate) of either green-only or green-then-red pulse pairs ($592\pm2$ ps separation time). Green-only illumination decreases the NV$^-$ population with power (see Fig.~\ref{A2}a),  matching results of Chen {\it et. al.} \cite{chen_spin_2015} in the low power regime. For the green-then-red results we plot the red-induced change in steady state NV$^-$ population to reveal the effects of the additional red pulses. As shown in Fig.~\ref{A2}b, the additional red illumination can cause either an increase or decrease in NV$^-$ population, depending on the green power. The amount of red-induced NV$^-$ population increase is maximized at low green and red powers, while increasing the red power beyond this point decreases population at the same rate regardless of green power. These results agree with recent literature \cite{hopper_near-infrared-assisted_2016,ji_charge_2016} and point to long-timescale internal dynamics that lead to qualitatively different behaviour than what we observe in the single-pulse regime. %Data such as that shown in Fig.~\ref{A2} does not sufficiently constrain the rates needed to model 

%Notes: Reduced Chi-squared for ionization to ground vs excited state

\bibliography{mybibliography2}

\end{document}